\documentstyle[11pt,amstex,amssymb,epsf]{article} 

\def\rect#1#2{{\vcenter{\vbox{\hrule height.3pt
            \hbox{\vrule width.3pt height#2truecm \kern#1truecm
            \vrule width.3pt}
            \hrule height.3pt}}}}

\def\finedim{\rule{0.25cm}{0.3cm}}

\newtheorem{theorem}{Theorem} 
\newtheorem{definition}[theorem]{Definition} 
\newtheorem{lemma}[theorem]{Lemma} 

\begin{document}
\addtolength{\baselineskip}{\baselineskip}
\title{Finite size scaling in three-dimensional bootstrap percolation}
\author{Rapha\"el Cerf and Emilio N.M. Cirillo\\
Universit\'e Paris Sud, Math\'ematique, \\
B\^atiment 425, 91405 Orsay Cedex, France}
\maketitle	

\vskip 2 cm
\begin{abstract}
\addtolength{\baselineskip}{\baselineskip}
We consider the problem of bootstrap percolation on a three dimensional
lattice and we study its finite size scaling behavior.
Bootstrap percolation is an example of Cellular Automata defined on the
$d$-dimensional lattice $\{1,2,...,L\}^d$ in which
each site can be empty or occupied by a single particle; in the
starting configuration each site is occupied with probability $p$, 
occupied sites remain occupied for ever, while empty sites are occupied
by a particle if at least $\ell$ among their $2d$ nearest neighbor sites are 
occupied.
When $d$ is fixed, the most interesting case is the one $\ell=d$: this
is a sort of threshold, in the sense that the critical probability
$p_c$ for the dynamics on the infinite lattice ${\Bbb Z}^d$
switches from zero to one when this limit is crossed.
Finite size effects in the three-dimensional case are already known
in the cases $\ell\le 2$: in this paper we discuss the case $\ell=3$
and we show that the finite size scaling function for this problem is
of the form $f(L)={\mathrm{const}}/\ln\ln L$. 
We prove a conjecture proposed by A.C.D. van Enter.
\end{abstract}


\newpage
\addtocounter{section}{+1}
Cellular Automata are dynamical systems defined on the $d$-dimensional 
lattice ${\Bbb Z}^d$ in which each site of the lattice is occupied 
by one of finitely many {\it types} at each time $t$. An 
updating rule is defined, which is {\it homogeneous}, all the sites
follow the same rule, and {\it local}, transitions are determined by the
configuration of types on a finite set of neighboring sites \cite{[U],[N]}.
\par
These models can be thought as interacting particle systems and
their connections with statistical mechanics models have been widely
studied in past years (see, for instance, \cite{[G],[TM],[V],[W]}). 
A particular example of cellular automata, known as {\it bootstrap 
percolation}, has been introduced in \cite{[CLR]} to model some
magnetic systems. More informations on the physical relevance of this
model are given in \cite{[KL],[tutti]}.
\par
In bootstrap percolation only two different types are associated to
each site: each site can be occupied or not by a particle. In the
starting configuration each site is independently occupied with probability $p$, 
occupied sites remain occupied for ever, while empty sites are occupied
by a particle if at least $\ell$ among their $2d$ nearest neighbor sites are 
occupied.
The object of primary interest is the probability $p_{\mathrm{full}}(\ell)$ 
that at the end of the dynamics, that is in the infinite time configuration, 
all the sites are occupied.
The basic question that has been addressed in physics literature is 
whether by 
changing the value of the parameter $p$ the system exhibits a sort of
phase transition, that is whether there exists a critical
value $p_c(\ell)\in [0,1]$ such that if $p\ge p_c(\ell)$ then 
$p_{\mathrm{full}}(\ell)=1$, otherwise $p_{\mathrm{full}}(\ell)< 1$.  
\par
Fixed the dimension $d$, the smaller $\ell$ the easier empty sites are 
occupied, hence one expects
that $p_c(\ell)$ is an increasing function of $\ell$. 
The first rigorous result on this topic is due to van Enter, who used
the idea of the Straley's argument \cite{[KL]} to prove
that in the case $d=2$ and $\ell=2$ the critical probability is equal to zero.
In \cite{[Sc1],[Sc2]} Schonmann has proved that $p_c(\ell)\in\{0,1\}$,
more precisely $p_c(\ell)=0$ if $\ell \le d$, otherwise $p_c(\ell)=1$;
these results suggest that the most peculiar case is $\ell=d$.
\par
Before these rigorous results the phase transition scenario in bootstrap
percolation models was not clear. The technique that had been used to measure
the critical probability $p_c(\ell)$ was the finite size scaling: a finite
volume estimate of the critical probability was found by means of Monte
Carlo simulations on a finite lattice $\Lambda_L=\{1,2,...L\}^d$, for instance 
the probability $p_L^{0.5}$ that
one half of the samples were completely filled at the end of the dynamics, 
and the critical value $p_c(\ell)$ was extrapolated
by means of a suitable scaling function $f(L)$. That is the expression   
\begin{equation}
p_{L}^{0.5}-p_c(\ell) \stackrel{L\to\infty}{\sim} f(L)
\label{fss}
\end{equation}
was supposed to be valid and Monte Carlo data were properly fitted by 
means of the function $f(L)$ (see \cite{[AA]} and references therein). 
\par
It is rather clear that the estimate of $p_c$ strongly depends on the choice
of the scaling function $f(L)$: the typical choice, when critical effects
in second order phase transition are studied, is 
$f(L)={\mathrm{const}}\times L^{-1/\nu}$ where $\nu$ is a suitable exponent.   
Actually this choice with $1/\nu=d$ is correct in the case $\ell=1$, while
estimations of $p_c(\ell)$ in the cases $\ell=2$ and $d\ge 2$
obtained by means of Monte Carlo data analyzed
through this function $f(L)$ did not fit in the rigorous scenario depicted
by Schonmann's results \cite{[AA]}: the problem is that  
the power law $L^{-1/\nu}$ approaches zero too quickly and must
be replaced by a slower function $f(L)={\mathrm{const}}\times (\ln L)^{-(d-1)}$
as suggested by the finite volume Aizenman and Lebowitz's results 
\cite{[AL],[LZ]}.
Indeed the analysis of old and new data performed via the correct
scaling function yields the correct estimate of the critical probability
\cite{[ASA],[EAD1],[EAD2]}.
\par
In \cite{[AL]} bootstrap percolation on finite lattices $\Lambda_L$ is
considered in the case $\ell=2$ and $d\ge 2$
and it is observed that if $p$ is kept fixed, then in the
limit $L\to\infty$ the probability $p_{\mathrm{full}}^{L,p}$  
to fill $\{1,...,L\}^d$ tends to one
whatever the value of $p$ is. But if $p\to 0$ together with $L\to \infty$,
then it is possible to find a particular regime in which the probability
to fill everything tends to zero. Indeed they prove that there exist
two constants $c_+>c_->0$ such that if $p\ge c_+ / (\ln L)^{d-1}$ then
$p_{\mathrm{full}}^{L,p}\to 1$ when $p\to 0$ and $L\to\infty$, while
if $p\le c_- / (\ln L)^{d-1}$ then in the same limit 
$p_{\mathrm{full}}^{L,p}\to 0$.  
\par
Let us focus on the case $d=3$: the choice $\ell=2$ is not the most
delicate one, indeed according to \cite{[Sc2]} even in the case $\ell=3$,
that is even in a situation in which it is more difficult to fill empty
sites, the critical probability is still zero. Hence one can guess 
that in the three-dimensional case if $\ell=3$
then the finite scaling function is no more the Aizenman-Lebowitz
one, but a function approaching zero more slowly. Our aim
is to study this case and to show that results similar to those in
\cite{[AL]}, and conjectured by A.C.D. van Enter,
can be proved if the scaling function is replaced 
by $f(L)={\mathrm{const}}/\ln\ln L$. 
This problem has been proposed in \cite{[Sc2]} as Problem 3.1; we notice
also that related problems have been discussed in \cite{[M],[Sc3]}.
\par
An interesting follow up would be the generalization of our
results to the $d$-dimensional case by considering 
arbitrary dimension $d$ and $\ell=d$.
In this case one expects 
$f(L)={\mathrm{const}}/(\ln\ln\; ...\;\ln L)$ with the logarithm applied
$d-1$ times.
\par
We next define the particular model of bootstrap percolation
that we are going to study and we introduce some notations.
Let us consider the lattice ${\Bbb Z}^3$ and the discrete time
variable $t=0,1,2,...$. To each site $x\in {\Bbb Z}^3$ 
we associate at each instant
of time $t$ a random variable $X_t(x)$ which takes values in
$\{0,1\}$; depending on $X_t(x)$ equals $0$ or $1$ we say that
the site $x$ is empty or occupied.   
We denote by $\Omega:=\{0,1\}^{{\Bbb Z}^3}$ the space of configurations
and by $X_t\in\Omega$ the configuration of the system at time $t$.
The initial configuration $X_0$ is chosen by occupying independently each site
of the lattice with probability $p$ (initial density). Then
the system evolves according to the following deterministic rules:
\begin{itemize}
\item
if $X_t(x)=1$, then $X_{t+1}(x)=1$ (1's are stable);
\item
if $X_t(x)=0$ and $x$ has at least three occupied sites
among its six nearest neighbors, then $X_{t+1}(x)=1$;
\item
$X_{t+1}(x)=0$ otherwise.
\end{itemize}
We omit in our notations the dependence of the process $X_t$ on the initial
density $p$; when the initial density will be different from $p$ it will be
clearly stated. 
\par
The primary object of interest for this problem is
the final configuration 
\begin{equation}
\label{eq:finale}
X:=\lim_{t\to\infty} X_t\;\;\;, 
\end{equation}
that is the configuration attained by the system when the dynamics stops. 
\par
Let us consider a subset $\Lambda\subset{\Bbb Z}^3$, we denote by 
$X_{\Lambda,t}$ the process restricted to $\Lambda$ with free boundary
conditions, that is without taking into account sites outside $\Lambda$.
In this case $X_{\Lambda,t}$, $X_{\Lambda,t}(x)$ and $X_{\Lambda}$ will
respectively denote the configuration at time $t$, the value of the
random variable at time $t$ and site $x$, and the final configuration.
\par
Given two arbitrary sets $\Lambda_1,\Lambda_2\subset{\Bbb Z}^3$ we will
consider the process $X_{\Lambda_1,t}^{\Lambda_2}$ restricted to
$\Lambda_1$ and with occupied sites in $\Lambda_2$
(the sites in ${\Bbb Z}^3\setminus (\Lambda_1\cup\Lambda_2)$ are
not taken into account). 
We will obviously omit $\Lambda_1$ in the notation 
if $\Lambda_1={\Bbb Z}^3$, while 
we will omit $\Lambda_2$ if $\Lambda_2=\emptyset$. 
\par
\begin{definition}
\label{def:spazzato}
Following \cite{[AL]} we say that a set $\Lambda\subset{\Bbb Z}^3$ is
{\it internally spanned} if it is entirely covered in the 
final configuration of the dynamics restricted to $\Lambda$, that is if
\begin{equation}
\label{eq:spazzato}
\forall x\in\Lambda\qquad
X_{\Lambda}(x)=1
\;\;\; .
\end{equation}
\end{definition}
\par
We have now introduced the bootstrap percolation
model in dimension $d=3$ and with parameter $\ell=3$; as we have already
stated, this is the most delicate three-dimensional
bootstrap percolation model. Indeed $\ell=3$ is the highest value of the 
parameter $\ell$ such that the critical probability in infinite volume
is equal to zero, that is
for each positive initial density $p$ the probability that the whole lattice 
will be completely occupied by particles at the end of the dynamics is 
equal to one.
\par
In this particular case we examine the question of finite size scaling,
that is, following \cite{[AL]}, we consider the process $X_{\Lambda_L,t}$
on the finite cube $\Lambda_L=\{1,...,L\}^3$ 
of size $L$ and we perform the limit
$L\to\infty$ and $p\to 0$. We prove that there exists a particular regime
in which the probability that the cube $\Lambda_L$ is internally
spanned is zero in the limit $L\to\infty$ and $p\to 0$. 
\begin{theorem}
\label{th:risultato}
Let us consider the cube $\Lambda_L$ of side length $L$
and the process $X_t$ with
initial density $p$; let us denote by $R(L,p)$ the probability
that the cube $\Lambda_L$ is internally spanned. There exist two 
constants $c_+>c_->0$ such that
\begin{itemize}
\item
$R(L,p)\longrightarrow 1$ if $(L,p)\to (\infty ,0)$ in the regime
$p>c_+/\ln\ln L$ 
\item
$R(L,p)\longrightarrow 0$ if $(L,p)\to (\infty ,0)$ in the regime
$p<c_-/\ln\ln L$ 
\end{itemize}
\end{theorem}

\par
We start to prove the first part of Theorem \ref{th:risultato}: 
this is the easy part of the theorem and its proof has already been 
sketched in \cite{[EAD1]}.
The idea of the proof relies on the notion of {\it critical length}:
suppose $p$ is small, if an occupied cube has size large
enough, then the probability to find on its faces two-dimensional occupied
square droplets so large to cover the whole faces of the cube (two-dimensional
critical droplets) is close to one. Obviously if $p$ goes to $0$,
then the size of the cube must diverge and this must happen fastly enough;
in this way one will find that the critical length is of order
$\exp({\mathrm{const}}/p)$.
Then, one estimates that in order to
have a high probability of finding a critical droplet, the size $L$
of the cube must be of order $\exp\exp({\mathrm{const}}/p)$.
\par
Now, we come to the proof: from results in 
\cite{[Sc2]} one easily
obtains that there exists a constant $c_1>0$ such that, given a cube
$\Lambda_l$, if $p\ge c_1/\ln l$ then there exists a constant $a_1>0$
such that
\begin{equation}
\label{eq:facile1}
P(\Lambda_l\;{\mathrm{covers}}\;{\Bbb Z}^3\; |\;
\Lambda_l\;{\mathrm{occupied\; at\;}} t=0)\ge
\prod\limits_{k=l}^{\infty} (1-{\mathrm{e}}^{-a_1 k})\;\;\; .
\end{equation}
We consider a large cube $\Lambda_L$ and we estimate $R(L,p)$:
\begin{displaymath}
R(L,p)\ge 
P({\mathrm{in}}\;\Lambda_L\;\exists\Lambda_l\;
{\mathrm{occupied\; at}}\; t=0,
\; \Lambda_l\;{\mathrm{covers}}\;{\Bbb Z}^3)=
\end{displaymath}
\begin{equation}
\label{eq:facile2}
P(\;\Lambda_l\;{\mathrm{covers}}\;{\Bbb Z}^3\;
|\;{\mathrm{in}}\;\Lambda_L\;\exists\Lambda_l\;
{\mathrm{occupied\; at}}\; t=0)
\; P({\mathrm{in}}\;\Lambda_L\;\exists\Lambda_l\;
{\mathrm{occupied\; at}}\; t=0)
\end{equation} 
Now, from equation (\ref{eq:facile1}) it follows that if 
$l\ge\exp(c_1/p)$ then the first factor in (\ref{eq:facile2})
tends to $1$ when $p$ goes to $0$. Then, we just have to prove that with
such an $l$ it is possible to choose $L$ such that the second factor
of (\ref{eq:facile2}) tends to $1$, as well. Indeed by partitioning
the cube $\Lambda_L$ in disjoint cubes of size $l$ one has 
\begin{equation} 
P({\mathrm{in}}\;\Lambda_L\;\exists\Lambda_l\;
{\mathrm{occupied\; at}}\; t=0)\ge
1-(1-p^{l^3})^{L/l}; 
\end{equation}
by choosing $c_+>3c_1>0$ and 
$L>\exp\exp(c_+/p)$, the right hand
term in the equation above tends to $1$ when $p$ goes to $0$.
This completes the proof of the first part of Theorem \ref{th:risultato}.
\par
Next we prove the second part of Theorem \ref{th:risultato} by finding
a suitable upper bound to the probability that a cube $\Lambda_L$ is
internally spanned.
First of all we give some more definitions: 
given a site $x\in{\Bbb Z}^3$ we denote by $(x_1,x_2,x_3)$ its
three coordinates and given a set $\Lambda\subset{\Bbb Z}^3$
we define its diameter
\begin{equation}
\label{eq:diametro}
d(\Lambda):=\sup\{|x_i-y_i|:\; x\in\Lambda,\; y\in\Lambda,\; i\in\{1,2,3\}\}
\;\;\; , 
\end{equation}
that is $d(\Lambda)$ is the side length of the minimal cube
surrounding the set $\Lambda$. 
We say that $\Lambda\subset{\Bbb Z}^3$ is a region of ${\Bbb Z}^3$ if
and only if it is nearest neighbors connected.
We note that if $\Lambda_1$ and 
$\Lambda_2$ are two regions of ${\Bbb Z}^3$ 
and $\Lambda_1\cup\Lambda_2$ is a region as well, then 
$d(\Lambda_1\cup\Lambda_2)\le d(\Lambda_1)+d(\Lambda_2)$.
\par
Now we adapt to our situation a key Lemma of \cite{[AL]} that
describes what happens at a smaller scale inside an internally 
spanned region.
\begin{lemma}
\label{lem:AL3d}
Let $\Lambda_1$ be a region of 
${\Bbb Z}^3$. If $\Lambda_1$ is internally spanned,
then for all $\kappa$ such that $1\le \kappa$ and 
$2\kappa+1\le d(\Lambda_1)$ there exists
at least a region $\Lambda_2$ included in $\Lambda_1$ which is
internally spanned and such
that $\kappa\le d(\Lambda_2)<2\kappa+1$.
\end{lemma}
\par{\it Proof.}
We build the configuration $X_{\Lambda_1}$ by means of the following 
algorithmic procedure.
Let ${\cal C}_0$ be the collection of the sites occupied at time zero. Suppose
we have built a collection of regions internally spanned ${\cal C}_n$, we
define a rule to build the collection ${\cal C}_{n+1}$:
\begin{itemize}
\item
If there exist two regions $A,B$ of ${\cal C}_n$ such that $A\cup B$  
is still a region, then we set
\begin{equation}
\label{eq:alg1}
{\cal C}_{n+1}:={\cal C}_n\cup\{A\cup B\}\setminus\{A,B\}\;\; ,
\end{equation}
that is ${\cal C}_{n+1}$ is obtained by replacing in ${\cal C}_n$ the
two elements $A$ and $B$ by $A\cup B$.
\item
If no such regions exist then we choose a site $x$ not
belonging to any set in ${\cal C}_n$ and having three neighbors in the
set $\bigcup\limits_{A\in{\cal C}_n}A$. We denote by $A_i$ with $1\le i\le r$
the $r$ regions of ${\cal C}_n$ containing a neighbor of $x$, and we
set
\begin{equation}
\label{eq:alg2}
{\cal C}_{n+1}:={\cal C}_n\cup
\left\{\bigcup\limits_{i=1}^r A_i\cup\{x\}\right\}
\setminus\left\{A_1,...,A_r\right\} \;\;\; .
\end{equation}
\item
If no such site $x$ exists, the algorithm stops.
\end{itemize}
Notice that for each $n$, the regions of ${\cal C}_n$ are internally 
spanned.
Since $\Lambda_1$ is internally spanned the procedure ends for some 
$m$ such that ${\cal C}_m=\{\Lambda_1\}$.
Moreover we have that 
$\max\{d(A):\; A\in{\cal C}_0\}=1$,
$\max\{d(A):\; A\in{\cal C}_m\}=d(\Lambda_1)$
and for any $n\le m-1$
\begin{equation}
\label{eq:alg3}
\max\{d(A):\; A\in{\cal C}_{n+1}\}\le 2\max\{d(A):\; A\in{\cal C}_n\}+1
\;\;\; . 
\end{equation}
Hence there exists $n$ such that:
\begin{equation}
\label{eq:alg4}
\kappa\le\max\{d(A):\; A\in{\cal C}_n\} < 2\kappa+1\;\;\; ,
\end{equation}
which means that in ${\cal C}_n$ there is an internally spanned region $A$
of diameter $d(A)$ such that $\kappa\le d(A)<2\kappa+1$. 
$\finedim$
\par
\begin{definition}
\label{def:crossed}
Let us consider a cube $\Lambda$ in ${\Bbb Z}^3$. We say that
$\Lambda$ is crossed, or that there is a crossing in $\Lambda$,
if and only if in the final configuration 
$X_{\Lambda}$ of the dynamics restricted to $\Lambda$
there is an occupied region joining
two opposite faces of the cube $\Lambda$.
\end{definition}
We note that for any region $A$ the following inclusion stands:
\begin{equation}
\label{eq:inclusione}
\{A\;{\mathrm{is\; internally\; spanned}}\}\subset
\{{\mathrm{the\; smallest\; cube\; surrounding\;}}A\;
{\mathrm{is\; crossed}}\}
\;\;\; .
\end{equation}
Hence, for any $L$ and any $\kappa$ such that $2\kappa+1 < L$, we have:
\begin{displaymath}
R(L,p)\le
P(\exists l,\; \kappa\le l<2\kappa+1,\;
\exists\Lambda_l\subset\Lambda_L,\;\Lambda_l
\;{\mathrm{is\; crossed}})\le 
\end{displaymath}
\begin{equation}
\label{eq:stima1}
(\kappa+1)\; L^3\max\limits_{\kappa\le l < 2\kappa+1} 
P(\Lambda_l\;{\mathrm{is\; crossed}})
\end{equation}
Thus,
\begin{equation}
\label{eq:stima2}
R(L,p)\le
L^3\min\limits_{1\le\kappa<(L-1)/2}
(\kappa+1)\; \max\limits_{\kappa\le l < 2\kappa+1} 
P(\Lambda_l\;{\mathrm{is\; crossed}})\;\; .
\end{equation}
\par
We have reduced the estimate of $R(L,p)$ to the estimate of the probability
that a cube $\Lambda_l$ is crossed and by symmetry we can consider the case
of a 
crossing along the first reticular direction (denoted by $e_1$ in the 
sequel):
\begin{equation}
\label{eq:simmetria}
P(\Lambda_l\;{\mathrm{is\; crossed}})\le
3\; P(\Lambda_l\;{\mathrm{is\; crossed\; along\;}} e_1)
\;\;\; .
\end{equation}
In order to estimate the second hand term of equation (\ref{eq:simmetria})
we reconduct to a two-dimensional situation by properly cutting
the cube $\Lambda_l$ in slices of thickness two and 
perpendicular to the first reticular direction.
For the sake of definiteness we suppose $\Lambda_l:=\{1,2,...,l\}^3$,
$l$ an even number and we define the slices  
\begin{equation}
\label{eq:affetto}
T_k:=\{x\in\Lambda_l:\; x_1=2k-1\;{\mathrm{or}}\; x_1=2k\}\,,
\;\;\;\;\;
1\le k\le \frac{l}{2}\;\;\; .
\end{equation}
We define a map $s$ associating to each site in a slice the only nearest
neighbor along the first reticular direction belonging to the
same slice:
\begin{equation}
\label{eq:map}
\forall x\in\Lambda_l\;\;\;\;\;
s(x):=\left\{
\begin{array}{ll}
(x_1+1,x_2,x_3)&\;\;\;\;\;{\mathrm{if}}\; x_1\;{\mathrm{is\; odd}}\\ 
(x_1-1,x_2,x_3)&\;\;\;\;\;{\mathrm{if}}\; x_1\;{\mathrm{is\; even}}\\ 
\end{array}
\right.\;\;\; .
\end{equation}
The process $X_{T_k,t}^{\Lambda_l\setminus T_k}$, restricted 
to the slice $T_k$ and with all the sites
in $\Lambda_l\setminus T_k$ occupied, dominates the original process
in the same slice:
\begin{equation}
\label{eq:dominalastriscia}
\forall k\in\{1,...,\frac{l}{2}\}\quad\forall x\in T_k\quad\forall t\ge 0
\qquad
X_t(x)\le X_{T_k,t}^{\Lambda_l\setminus T_k}(x)
\;\;\; .
\end{equation}
In each slice $T_k$ for $k$ in $\{1,...,l/2\}$ we define a 
new process $Y_t^k$.
\begin{definition}
\label{def:processostriscia}
We consider all the sites in $\Lambda_l\setminus T_k$ occupied and
define the process $Y_t^k$ on $\{0,1\}^{T_k}$ as follows: 
for any $x$ in $T_k$ 
\begin{itemize}
\item
$Y_0^k(x)=\max (X_{T_k,0}^{\Lambda_l\setminus T_k}(x),
X_{T_k,0}^{\Lambda_l\setminus T_k}(s(x)))=
\max (X_0(x),X_0(s(x)))$
\item   
if $Y_t^k(x)=1$ then $Y_{t+1}^k(x)=1$
\item
if $Y_t^k(x)=0$ and $x$ has at least $3$ occupied sites among its $6$   
nearest neighbors, then $Y_{t+1}^k(x)=1$ and $Y_{t+1}^k(s(x))=1$ 
\item
$Y_{t+1}^k(x)=0$ otherwise
\end{itemize}
\end{definition} 
The mechanism to build $Y_{t+1}^k(x)$ is the one used
for $X_{T_k,{t+1}}^{\Lambda_l\setminus T_k}$ followed by an additional
step increasing the configuration.
This mechanism ensures that for all $t$ and any $x$ in $T_k$ one has
$Y_t^k(x)=Y_t^k(s(x))$.
\par
Finally we introduce a two-dimensional process that will be used
to estimate (\ref{eq:simmetria}).
\begin{definition}
\label{def:processo2d}
Let us associate to each slice $T_k$ for $k$ in $\{1,...,l/2\}$
a two-dimensional $l\times l$ square $Q_l^k:=\{1,2,...,l\}^2$; then 
on each square $Q_l^k$ we define a process $Z_{Q_l^k,t}$ by
\begin{equation}
\label{eq:processo2d}
\forall (x_2,x_3)\in Q_l^k
\;\;\;\;\;
Z_{Q_l^k,t}(x_2,x_3):=Y_t^k(2k-1,x_2,x_3)=Y_t^k(2k,x_2,x_3)
\;\;\; .
\end{equation} 
\end{definition}
The processes $Z_{Q_l^k,t}$, $1\le k\le l/2$, are independent
and they are two-dimensional bootstrap percolation processes with
parameter $\ell=2$ and initial density $q=2p-p^2$. 
Furthermore these processes dominate the original process on the
slices in the sense:   
\begin{equation}
\label{eq:2ddomina3d}
\forall k\in\{1,...,\frac{l}{2}\}\quad\forall x=(x_1,x_2,x_3)\in T_k\quad\forall t\ge 0
\qquad
X_{T_k,t}(x)\le Z_{Q_l^k,t}(x_2,x_3)
\;\;\; .
\end{equation}
\par
The two-dimensional processes can be used to estimate 
the probability that a cubic region is crossed: we consider the 
$(l/2)\times l\times l$ parallelepiped ${\cal P}_l$  
obtained by collecting the $l/2$ squares $Q_l^k$ 
and we denote by $Z_{{\cal P}_l}$ the configuration on ${\cal P}_l$
defined as follows:
\begin{equation}
\label{eq:confsucalP}
\forall x_1\in\{1,...,\frac{l}{2}\}\;\forall x_2,x_3\in\{1,...,l\}
\;\;\;\;\;
Z_{{\cal P}_l}(x_1,x_2,x_3):=Z_{Q_l^{x_1}}(x_2,x_3)
\;\;\; ,
\end{equation}
where $Z_{Q_l^k}$ with $k$ in $\{1,...,l/2\}$ 
is the final configuration of the process $Z_{Q_l^k,t}$.
Finally we have the way to estimate (\ref{eq:simmetria}):
\begin{equation}
\label{eq:stima3}
P(\Lambda_l\;{\mathrm{is\; crossed\; along\;}} e_1)
\le
P({\mathrm{in}}\; Z_{{\cal P}_l}\;{\mathrm{there\; is\; a\; crossing\;
along\;}} e_1)\;\;\; .
\end{equation} 
\par
Now we consider the two-dimensional bootstrap percolation with parameter
$\ell=2$ and initial density $q$. We define as well the concept of
``being internally spanned" and we denote by $S(l)$ a square of side length
$l$. We recall that the final configuration, for such a process,
is a union of separated rectangular regions and we state a few results:
\begin{lemma}[Aizenman - Lebowitz \cite{[AL]}]
\label{lem:AL2d}
For all $\kappa\ge 1$, a necessary condition for $S(l)$ to be
internally spanned, where $\kappa\le l$, is
that it contains at least one rectangular region whose maximal side length
is in the interval $[\kappa,2\kappa+1]$ which is also internally
spanned.
\end{lemma}
Proof of Lemma \ref{lem:AL2d} is given in \cite{[AL]}.
\begin{lemma}
\label{lem:2lemma2d}
Let $A$ be a rectangular region of side lengths $l_1$ and $l_2$; suppose
$l_1\le l_2$. For $q$ small enough one has
\begin{equation}
\label{eq:2lemma2d}
P(A\;{\mathrm{is\; internally\; spanned}})\le
(4l_2 q)^{l_2/2} 
\;\;\; .
\end{equation}
\end{lemma}
\par{\it Proof.} 
If $A$ is partitioned in $l_2/2$ disjoint slabs of width $2$, a necessary
condition for $A$ to be internally spanned is that each slab contains
initially an occupied site. Hence:
\begin{displaymath}
P(A\;{\mathrm{is\; internally\; spanned}})
\le\left(1-(1-q)^{2l_1}\right)^{l_2/2}\le
\end{displaymath}
\begin{displaymath}
\exp\left(\frac{l_2}{2}\;\ln\left(1-\exp(2l_2\;\ln(1-q))\right)\right)\le
\end{displaymath}
\begin{equation}
\exp\left(\frac{l_2}{2}\;\ln\left(-2l_2\;\ln(1-q)\right)\right)
\end{equation}
For $q$ small enough, $\ln(1-q)\ge -2q$, whence
\begin{equation}
P(A\;{\mathrm{is\; internally\; spanned}})\le
\exp\left(\frac{l_2}{2}\;\ln(4l_2 q)\right)=(4l_2 q)^{l_2/2}
\;\;\; .
\;\;\;\finedim
\end{equation}
\begin{lemma}
\label{lem:3lemma2d}
For any $l$ and  any $\kappa\le (l-1)/2$ 
let ${\cal E}$ be the event: $S(l)$ contains a rectangular region internally
spanned whose maximal side length belongs to the interval 
$[\kappa,2\kappa+1]$.
For $q$ small enough one has
\begin{equation}
\label{eq:3lemma2d}
P({\cal E})\le
l^2\; (2\kappa+1)^2\;\exp\left( -\frac{\kappa}{2}\;\exp (-4(2\kappa+1)q)
\right)
\;\;\; .
\end{equation}
\end{lemma}
\par{\it Proof.} 
We suppose $q$ small enough to have $\ln(1-q)\ge -2q$ and
we bound the probability of the event ${\cal E}$ as follows.
\begin{displaymath}
P({\cal E})
\le l^2\;(2\kappa+1)^2\;\max\limits_{\kappa\le l_2\le 2\kappa+1}
\exp\left(\frac{l_2}{2}\;\ln(1-\exp(2l_2\;\ln(1-q)))\right)\le
\end{displaymath}
\begin{displaymath}
l^2\;(2\kappa+1)^2\; \exp\left(\frac{\kappa}{2}\;
\ln(1-\exp(2(2\kappa+1)\;\ln(1-q)))\right)\le
\end{displaymath}  
\begin{displaymath}
l^2\;(2\kappa+1)^2\; \exp\left(\frac{\kappa}{2}\;
\ln(1-\exp(-4(2\kappa+1)q))\right)\le
\end{displaymath}
\begin{equation}
l^2\;(2\kappa+1)^2\; \exp\left(-\frac{\kappa}{2}\;
\exp(-4(2\kappa+1)q)\right) 
\;\;\; .\;\;\;\finedim
\end{equation}
\par
Now we come back to the proof of the upper bound. Let us consider $\alpha>0$, 
we notice that 
$p\le q\le 2p$.  
We denote
by $\underline 1$ a configuration in a square $Q_l^h$ with all the sites 
occupied.
We still increase the configuration $Z_{Q_l^k}$ by setting
$Z_{Q_l^k}=\underline 1$
in case $Z_{Q_l^k}$ contains at least one rectangular region of
maximal side length larger than $\alpha/q$. 
Supposing $l\ge\alpha/q$ and $q$ small enough so that $\alpha/q>3$, 
by applying Lemma \ref{lem:3lemma2d} with $\kappa=\alpha/3q$ one has
\begin{displaymath}
P(Z_{Q_l^k}=\underline 1)\le
l^2\;\left(\frac{2\alpha}{3q}+1\right)^2\;\exp\left(-\frac{\alpha}{6q}\;
\exp\left(-4\left(\frac{2\alpha}{3q}+1\right)q\right)\right)\le
\end{displaymath}
\begin{equation}
\label{eq:stima4}
\frac{l^2\alpha^2}{q^2}\;\exp\left(-\frac{\alpha}{6q}\;
\exp(-4\alpha)\right)
\end{equation}
We suppose that $\alpha$ is small enough to have $\exp(-4\alpha)\ge 1/2$;
thus,
\begin{equation}
\label{eq:stima4'}
P(Z_{Q_l^k}=\underline 1)\le
\frac{l^2\alpha^2}{q^2}\;\exp\left(-\frac{\alpha}{12q}\right)
\;\;\; .
\end{equation}
\par
Let $M$ be the number of indices $k$ such that 
one has $Z_{Q_l^k}=\underline 1$ and let 
$k(1),...,k(M)$ be these indices arranged in increasing order. 
Let ${\cal E}_1$ be the event $\{$there is a crossing 
along $e_1$ in $Z_{{\cal P}_l}\}$. We decompose this event as follows:
\begin{equation}
\label{eq:decomposition}
P({\cal E}_1)=
P({\cal E}_1,M=0)+
\sum\limits_{m=1}^{l/2}\;\sum\limits_{i_1<\dots<i_m}
P({\cal E}_1,\; M=m, k(1)=i_1,....,\; k(m)=i_m)\;\;\; .
\end{equation} 
Let $i<j$ be two indices in $\{1,...,l/2\}$; 
by ${\cal E}(i,j)$ we denote the following event: there exists a sequence
of $H$ disjoint rectangular regions $(R_h, \,1\le h\le H)$   
in $Z_{{\cal P}_l}$ such that
\begin{itemize}
\item
$R_1$ is included in $Q_l^i$, $R_H$ is included in $Q_l^j$;
\item
the regions $R_h$ with $2\le h\le H-1$ are included 
in $\bigcup\limits_{i<h<j}Q_l^h$; 
\item
the maximal side length of all these regions
is strictly less than $\alpha/q$: 
we denote by $r_h$ the maximal side length of $R_h$;
\item
for each $h$, $1\le h\le H-1$, a site of $R_h$ is the neighbor of a site
of $R_{h+1}$;
\item
all the sites of these regions are occupied in $Z_{{\cal P}_l}$.
\end{itemize}
We remark that $H$ is free, however it has to be larger
than $j-i+1$. Moreover the
sequence of rectangles $(R_h,\,2\leq h\leq H-1)$ can go back 
and forth between the squares $Q_l^{i+1}$ and $Q_l^{j-1}$.
We make the convention that
for any $i$ the events ${\cal E}(i,i)$ and ${\cal E}(i,i-1)$
are the full events.
We have the following estimate:
\begin{displaymath}
P({\cal E}_1,\; M=m, k(1)=i_1,....,\; k(m)=i_m)
\le \; P(M=m,\;k(1)=i_1,....,\; k(m)=i_m,\; 
\end{displaymath}
\begin{displaymath}
{\cal E}(1,i_1-1),\; 
{\cal E}(i_1+1,i_2-1),...,\; {\cal E}(i_{m-1}+1,i_m-1),\; 
{\cal E}(i_m+1,\frac{l}{2}))=
\end{displaymath}
\begin{displaymath}
\prod\limits_{h=1}^{m+1}
P({\cal E}(i_{h-1}+1,i_h-1))\; P(Z_{Q_l^{i_1}}=\underline 1)\cdot\cdot\cdot
P(Z_{Q_l^{i_m}}=\underline 1)\le
\end{displaymath}
\begin{equation}
\label{eq:stima7} 
\left(\frac{l^2\alpha^2}{q^2}\;\exp\left(-\frac{\alpha}{12q}\right)
\right)^m
\;\prod\limits_{h=1}^{m+1}
P({\cal E}(i_{h-1}+1,i_h-1))
\end{equation}
where we have set $i_0=0\;$ and $i_{m+1}=\frac{l}{2}+1$.
\par\noindent
In order to estimate $P({\cal E}(i,j))$ with $i<j$, we consider a fixed 
sequence 
$r_1,...,r_H$ in $\{1,...,\alpha/q-1\}$; the
number of sequences $R_1,...,R_H$ of rectangles with maximal
sides $r_1,...,r_H$ and satisfying the above requirements is smaller than
\begin{equation}
\label{eq:rettangoli}
l^2r_1\times r_1^2r_2\times r_2^2r_3\times ... \times r_{H-1}^2r_H\le 
l^2\; (r_1 r_2\cdot\cdot\cdot r_H)^3
\;\;\; .
\end{equation}
Notice that several rectangles of the sequence 
$R_1,...,R_H$ can belong to the same slice. 
However the rectangles are disjoint and the events that they are 
internally spanned depend only on the dynamics restricted to the
rectangles, hence these events are independent, so that
\begin{displaymath}
P(R_1,...,R_H\;{\mathrm{are\; occupied\; in\;}} Z_{{\cal P}_l})\le
P(R_1,...,R_H\;{\mathrm{internally\; spanned}})\le
\end{displaymath}
\begin{equation}
\label{eq:ancorarettangoli}
P(R_1\;{\mathrm{internally\; spanned}})
\cdot\cdot\cdot
P(R_H\;{\mathrm{internally\; spanned}})\le
(4r_1q)^{r_1/2}\cdot\cdot\cdot (4r_Hq)^{r_H/2},\;\;
\end{equation}
where in the last inequality we have used 
the Lemma \ref{lem:2lemma2d}. Thus
\begin{displaymath}
P({\cal E}(i,j))=
\sum\limits_{H\ge j-i+1}
P(\exists R_1,...,R_H\;{\mathrm{realizing}}\; {\cal E}(i,j))\le
\end{displaymath}
\begin{displaymath}
\sum\limits_{H\ge j-i+1} \sum\limits_{r_1,...,r_H<\alpha/q}
l^2\;(r_1\cdot\cdot\cdot r_H)^3\; (4r_1q)^{r_1/2}\cdot\cdot\cdot 
(4r_Hq)^{r_H/2}=
\end{displaymath}
\begin{equation}
\label{eq:stima8}
\sum\limits_{H\ge j-i+1} l^2\; \left(
\sum\limits_{1\le r<\alpha/q}r^3\;(4rq)^{r/2}\right)^H
\end{equation}
We estimate the sum $\sum\limits_{1\le r<\alpha/q}
r^3\;(4rq)^{r/2}$ as follows:
\begin{displaymath}
\sum\limits_{1\le r<\alpha/q}r^3\;(4rq)^{r/2}=
\sum\limits_{1\le r<8}
r^3\;(4rq)^{r/2}\;
+\;
\sum\limits_{9\le r<\alpha/q}
r^3\;(4rq)^{r/2}\le
\end{displaymath}
\begin{equation}
\label{stima8.0}
8^3(32q)^{1/2}+\left(\frac{\alpha}{q}\right)^4\;
\max\limits_{9\le r<\alpha/q}(4rq)^{r/2}
\;\;\; .
\end{equation}
Let $f(r)=(4rq)^{r/2}$; for $\alpha$ small $f(r)$ is decreasing
on $[9,\alpha/q[$, whence
\begin{equation}
\sum\limits_{1\le r<\alpha/q}r^3\;(4rq)^{r/2}
\le
8^3 (32q)^{1/2}+\alpha^4 36^{9/2} q^{1/2} \le b_0 q^{1/2} 
\label{eq:stima8.1}
\end{equation}
where $b_0$ is a constant not depending on $\alpha$. Thus,
\begin{equation}
P({\cal E}(i,j))\le
\sum\limits_{H\ge j-i+1}l^2(b_0\sqrt q)^H
\label{stima8.2}
\end{equation}
Finally, for $q$ small enough, so that $b_0\sqrt q<1/2$, one has
\begin{equation}
P({\cal E}(i,j))\le
2l^2(b_0\sqrt q)^{j-i+1}=
2l^2\exp((j-i+1)(\ln b_0+\frac{1}{2}\ln q))
\;\;\; .
\label{eq:stima8.3}
\end{equation}
Coming back to inequality (\ref{eq:stima7}) 
\begin{displaymath}
P({\cal E}_1,\; M=m, k(1)=i_1,....,\; k(m)=i_m)
\le
\end{displaymath}
\begin{equation}
\label{eq:stima12}
\left(\frac{l^2\alpha^2}{q^2}\exp\left(-\frac{\alpha}{12q}\right)\right)
^m (2l^2)^{m+1} \exp\left[\left(\frac{l}{2}-m\right)
\left(\ln b_0+\frac{1}{2}\ln q\right)\right]
\end{equation}
and
\begin{displaymath}
P({\cal E}_1)=
P({\cal E}_1,M=0)+
\sum\limits_{m=1}^{l/2}\;\sum\limits_{i_1<\dots<i_m}
P({\cal E}_1,\; M=m, k(1)=i_1,....,\; k(m)=i_m)\le
\end{displaymath}
\begin{displaymath}
\sum\limits_{m=0}^{l/2}\left(\frac{l}{2}\right)^m
\left(\frac{l^2\alpha^2}{q^2}\exp\left(-\frac{\alpha}{12q}\right)\right)
^m (2l^2)^{m+1} \exp\left[\left(\frac{l}{2}-m\right)
\left(\ln b_0+\frac{1}{2}\ln q\right)\right]
\le
\end{displaymath}
\begin{equation}
\label{eq:stima11}
2l^2 \exp\left[\frac{l}{2}\left(\ln b_0+\frac{1}{2}\ln q\right)\right]
\sum\limits_{m=0}^{l/2}
\left[\frac{l^5\alpha^2}{q^2}
\exp\left(-\frac{\alpha}{12q}\right)\exp\left[
-\left(\ln b_0+\frac{1}{2}\ln q\right)\right]\right]^m\,.
\end{equation}
This is our estimate of the probability of a crossing
in the case $l\ge\alpha/q$.
On the other hand, in the case $l< \alpha/q$
we estimate the probability of a crossing by considering
directly the event ${\cal E}(1,l/2)$:
\begin{equation}
\label{eq:stima16}
P(\Lambda_l\;{\mathrm{is\; crossed\; along}}\; e_1)\le 
P({\cal E}(1,\frac{l}{2}))\le
2l^2\exp\left[\frac{l}{2}\left(\ln b_0 + 
\frac{1}{2}\ln q\right)\right]\,.
\;\;\;  
\end{equation}
We note that it would have not been possible to use the same strategy in the
case $l\ge \alpha/q$ because the probability of having a very large 
internally spanned rectangle does not vanish.
\par
Supposing that $l\le\exp(\alpha/120q)$ one has
\begin{equation}
\label{eq:stima16.1}
\frac{l^5\alpha^2}{q^2}
\exp\left(-\frac{\alpha}{12q}\right)\exp\left[
-\left(\ln b_0+\frac{1}{2}\ln q\right)\right]
\le
\frac{\alpha^2}{q^2}\exp\left(
-\frac{\alpha}{24q}-\ln b_0-\frac{1}{2}\ln q\right)
\;\;\; . 
\end{equation}
Now, if $q$ is sufficiently small so that the right hand term is
smaller than $1$ and $\ln b_0\le -(1/4)\ln q$, under the hypothesis
$2\le l\le\exp(\alpha/120q)$, from (\ref{eq:simmetria}), (\ref{eq:stima11}), 
(\ref{eq:stima16}) and (\ref{eq:stima16.1}) one has 
\begin{equation}
\label{eq:stima17}
P(\Lambda_l\;{\mathrm{is\; crossed}})\le 
6l^2\exp\left[\frac{l}{2}
\left(\ln b_0 +\frac{1}{2}\ln q\right)\right]
\left(\frac{l}{2}+1\right)
\le
6l^3\exp\left(\frac{l}{8}\ln q\right)
\;\;\; .
\end{equation}
Hence, there exists $\alpha >0$ such that for $p$ sufficiently small
and $2\le l\le\exp(\alpha/240p)$
\begin{equation}
\label{eq:stima18}
P(\Lambda_l\;{\mathrm{is\; crossed}})\le 
6l^3\exp\left(\frac{l}{8}\ln (2p)\right)
\;\;\; .
\end{equation}
\par
Finally, we can use equations (\ref{eq:stima2}) and (\ref{eq:stima18})
to estimate the probability $R(L,p)$ and to complete the proof
of Theorem \ref{th:risultato}. First we consider the case
$L\le \exp(\alpha_0/p)$ with $\alpha_0=\alpha/240$ and we write  
\begin{equation}
\label{eq:stimafinale1}
R(L,p)\le
L^3\min\limits_{1\le\kappa<(L-1)/2}
(\kappa+1)\times 6(2\kappa+1)^3\exp\left(\frac{\kappa}{8}\ln(2p)\right)
\le
6L^7\exp\left(\frac{L}{24}\ln(2p)\right)
\end{equation}
and we remark that the right hand term goes to zero in the limit
$p\to 0$ and $L\to\infty$.
On the other hand, in the case $L>\exp(\alpha_0/p)$ one can restrict
the minimum to $2\kappa+1\le \exp(\alpha_0/p)$ and write 
\begin{equation}
\label{eq:stimafinale2}
R(L,p)\le
6L^3
\exp\left(\frac{4\alpha_0}{p}\right)
\exp\left[\frac{1}{24}\exp\left(\frac{\alpha_0}{p}\right)\ln(2p)\right]
\;\;\; .
\end{equation}
From the estimate above it is clear that there exists a positive constant
$c_-$ such that if $L$ is less than $\exp\exp(c_-/p)$ then 
$R(L,p)$ goes to $0$ in the limit
$L\to\infty$ and $p\to 0$. This
completes the proof of Theorem \ref{th:risultato}.
$\finedim$

\vskip 2 cm
\par\noindent
{\Large\bf Acknowledgements}
\vskip 0.3 cm
\par\noindent 
The authors express their thanks to A.C.D. van Enter who suggested the topic
of this paper and to R.H. Schonmann for useful suggestions and comments.


\newpage
\begin{thebibliography}{99}
\bibitem{[AA]}
J. Adler, A. Aharony (1988).
Diffusion percolation: I. Infinite time limit and bootstrap 
percolation.
{\sl J. Phys. A: Math. Gen.} {\bf 21}, 1387.

\bibitem{[ASA]}
J. Adler, D. Stauffer, A. Aharony (1989).
Comparison of bootstrap percolation models.
{\sl J. Phys. A: Math. Gen.} {\bf 22}, L297.

\bibitem{[AL]}
M. Aizenman, J.L. Lebowitz (1998).
Metastability effects in bootstrap percolation.
{\sl J. Phys. A: Math. Gen.} {\bf 21}, 3801.

\bibitem{[tutti]}
N.S. Branco, R.R. Dos Santos, S.L.A. de Queiroz,
{\sl J. Phys. C} {\bf 17}, L373 (1984);
M.A. Khan, H. Gould, J. Chalupa, 
{\sl J. Phys. C} {\bf 18}, L233 (1985);
N.S. Branco, S.L.A. de Queiroz, R.R. Dos Santos, 
{\sl J. Phys. C} {\bf 19}, 1909 (1986).

\bibitem{[CLR]}
J. Chalupa, P.L. Leath, G.R. Reich (1979).
Bootstrap percolation on a Bethe lattice. 
{\sl J. Phys. C: Solid State Physics} {\bf 12}, L31.

\bibitem{[E]}
A.C.D. van Enter (1987).
Proof of Straley's argument for Bootstrap Percolation.
{\sl J. Statist. Phys.} {\bf 48}, 943.

  
\bibitem{[EAD1]}
A.C.D. van Enter, J. Adler, J.A.M.S. Duarte (1990).
Finite-size effects for some bootstrap percolation models.
{\sl J. Statist. Phys.} {\bf 60}, 323.

\bibitem{[EAD2]}
A.C.D. van Enter, J. Adler, J.A.M.S. Duarte (1991).
Addendum: Finite-size effects for some bootstrap percolation models.
{\sl J. Statist. Phys.} {\bf 62}, 505.

\bibitem{[G]}
D. Griffeath (1998).
Cyclic random competition: a case history in experimental mathematics.
{\sl Notices of the American Mathematical Society} {\bf 35}, 1472.

\bibitem{[KL]}
P.M. Kogut, P.L. Leath (1981).
{\sl J. Phys. C: Solid State Physics} {\bf 14}, 3187.

\bibitem{[LZ]}
R. Le Normand, C. Zarcone (1984).
``Kinetics of Aggregation and Gelation", eds. F. Family and D.P. Landau  
(Elsevier, Amsterdam).

\bibitem{[M]}
T.S. Mountford (1995).
Critical length for semi-oriented bootstrap percolation.
{\sl Stochastic Process. Appl.} {\bf 56}, 185-205.


\bibitem{[N]}
J. von Neumann (1966).
Theory of self-reproducing automata.
Univ. of Illinois Press, Urbana.

\bibitem{[Sc1]}
R.H. Schonmann (1990).
Critical Points of two-dimensional bootstrap percolation-like cellular
automata.
{\sl J. Statist. Phys.} {\bf 58}, 1239.

\bibitem{[Sc3]}
R.H. Schonmann (1990).
Finite size scaling behavior of a biased majority rule cellular
automaton.
{\sl Phys. A} {\bf 167}, 619-627.

\bibitem{[Sc2]}
R.H. Schonmann (1992).
On the behavior of some cellular automata related to bootstrap
percolation.
{\sl Ann. Probab.} {\bf 20}, 174.

\bibitem{[TM]}
T. Toffoli, N. Margolus (1987).
Cellular automata machines. A new environment for modeling.
MIT Press.

\bibitem{[U]}
S. Ulam (1950).
Random processes and transformations.
Proc. Internat. Congr. Math., 264-275.

\bibitem{[V]}
G.Y. Vichniac (1984).
Simulating physics with cellular automata.
{\sl Phys. D} {\bf 10}, 96.

\bibitem{[W]}
S. Wolfram (1983).
Statistical mechanics of cellular automata.
{\sl Rev. Modern Phys.} {\bf 55}, 601 (1983).
S. Wolfram (1986).
Theory and applications of cellular automata.
World Scientific, Singapore.

\end{thebibliography}
\end{document}